# Measuring obscured complex OAM spectrum using Stokes fluctuations

Tushar Sarkar,[1] Reajmina Parvin,[2] Maruthi M. Brundavanam,[2] Rakesh Kumar Singh,[1,*]

[1] Department of Physics, Indian Institute of Technology (Banaras Hindu University), Varanasi, 221005, Uttar Pradesh, India
[2] Department of Physics, Indian Institute of Technology Kharagpur, Kharagpur – 721302, India
*Corresponding author: krakeshsingh.phy@iitbhu.ac.in

## Abstract
A novel technique to measure the orbital angular momentum (OAM) spectrum of the beam obscured by a complex random media is proposed and experimentally demonstrated. This is realized by measuring the complex correlation polarization function (CPCF) with the help of a two-point Stokes fluctuations correlation of the random light. The OAM spectrum analysis is implemented by projecting a complex field of the CPCF into spiral harmonics. A detailed theoretical framework is developed to measure complex amplitude and to decompose different integer OAM states of the obscured beam. The developed theoretical framework is verified by numerical simulation and tested by experimental demonstration of OAM mode compositions of the fractional optical vortex (FOV) beam.

## Introduction
A light beam carrying helical mode has attracted tremendous interest in recent years owing to its unique features, exhibiting a spiral phase structure[1-5]. This is inscribed by a spiral phase factor $e^{il\varphi}$, where $\varphi$ denotes azimuthal angle and $l$ the topological charge (TC) denotes an explicit amount of OAM equal to $l\hbar$ per photon[1-4]. The beam with integer TC is referred to as an integer vortex beam. The integer vortex beam reveals a null in the heart of the amplitude due to the point singularity around the spiral phase[1-3]. On the other hand, a beam with non-integer TC ($l$) is referred to as a FOV. The phase structure of the FOV contains a single mixed screw edge dislocation and the intensity distribution is no longer circularly symmetric[1-5]. The FOV could be deemed as multiplexing integer OAM beams with different intensity weights[2,3]. The composition of OAM modes and FOV show prominent applications in quantum digital spiral imaging[4], free-space optical communication, and sensing[1-3]. The measurement of the OAM mode spectrum is also central to others applications such as cryptography, unconventional interferometer, etc.

Several techniques have been proposed over the years to detect OAM modes compositions. These techniques include mode decomposition using digital hologram[6,7], probing the OAM spectrum using the tilted lens[8], analysis of OAM spectrum using the holographic approach, and spatial light modulator[9], measuring OAM spectrum by digital analysis of interference pattern[10]. In addition, the OAM discrimination method was developed to measure FOV and identify the difference between intrinsic OAM and total OAM for fractional OAM states[11]. A method was proposed to quantitatively measure non-integer OAM with a cylindrical lens and a camera[12]. Recently, a method was proposed for precision measurement of fractional OAM based on a two-dimensional multifocal array consisting of different integer vortices[13]. The OAM sorting methods based on the ray optics coordinate transformation have also been developed[14,15]. However, these methods consider propagation in free space or homogeneous media.

On the other hand, measuring the OAM spectrum of the target obscured by the complex random media is a challenging task and yet a highly practical problem owing to the scrambling of the spatial structure of the light[5,16]. When the beam propagates through a scattering media, the inhomogeneity in the optical path lengths scrambles the target beam[5,16,20]. The scrambled light plays a detrimental role in optical communication, remote sensing, and wireless communication[17,18]. Therefore, previously mentioned techniques are not capable to measure the OAM spectrum of the obscured beam[5-15]. In a separate investigation, sorting of spatially incoherent optical vortex mode is implemented using two-phase masks such as for transformation and the correction in the two-shot intensity measurements[19].

This article proposes a new method to measure the OAM spectrum of the beam from higher-order Stokes Parameters (SPs) fluctuation correlations of the randomly light. A detailed theoretical framework is established to measure the OAM spectrum using SPs fluctuations correlations. The correlations of the fluctuations of all SPs of the random light are evaluated which imparts a 4 x 4 correlation matrix with a total of sixteen elements. Out of these sixteen elements, only four elements help to extract CPCF from spatially fluctuating random light. Hereinafter, the complex field of the CPCF projection over the helical basis is applied to examine the composition of the OAM spectrum. The proposed analysis technique is free from using the specialized mask and additionally non-interferometric, iteration-free, and free from the pre-calibration requirement of the scattering medium[17]. Due to these unique features, our experimental technique offers high flexibility and robustness. The proposed theoretical framework is verified by simulation results and also confirmed by experimental results. The application of our technique is demonstrated in the recovery of the quantitative information of the FOV obscured by the random scattering medium and measuring decomposition of the FOV in different integer OAM states. The detailed theoretical explanation and the corresponding experimental test are discussed below.

## Methodology

Consider a transversely polarized beam with orthogonal polarization states x and y. The complex field at the transverse plane z=0 is expressed as

$$E(\hat{r}) = A_0(\hat{r}) \exp(i\Psi) \hat{e}_x + A\hat{e}_y, \tag{1}$$

where $\hat{e}_x$ and $\hat{e}_y$ are horizontal and vertical polarization states of the light respectively and $\hat{r}$ represents spatial position vector at the transverse plane. $A_0(\hat{r})$, $A$ denotes the amplitude of the target beam and non vortex beam respectively, $\Psi$ is the phase structure of the target beam.

Now the polarized light represented by Eq. (1) propagates through a random scattering medium and further travels down to the observation plane located in the far-field. A conceptual representation of the propagation and generation of the random field is shown in Fig. 1. The random scattering leads to a speckle pattern at the detector which is represented as.

$$E(r) = \mathscr{F}\big(E(\hat{r})e^{i\delta(\hat{r})}\big) = \hat{e}_x E_x(r) + \hat{e}_y E_y(r), \tag{2}$$

where $E(r)$ indicates the scattered field at the far-field, $\delta(\hat{r})$ denotes the spatial random phase introduced by the non-birefringent random scatterer, $r$ represents the spatial position vector at the observation plane, and $\mathscr{F}$ denotes two dimensional Fourier transform.

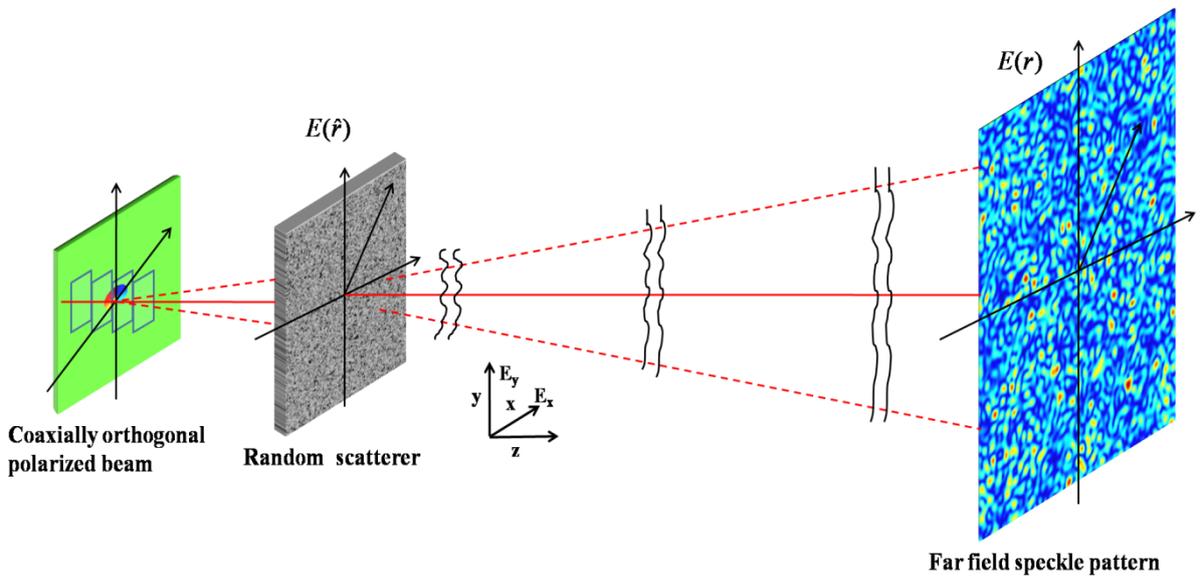

**Far field speckle pattern**

Figure 1. Conceptual representation of the generation of the Speckle pattern by orthogonally polarized light fields.

The fluctuations between the SPs of the speckle field provide the complex polarization correlation function which carries the information of the target beam. The SPs can be expressed in terms of four Pauli spin matrices as[20, 21]

$$S_n(r) = E^\dagger(r)\sigma^n E(r), \qquad (n = 0, 1, 2, 3), \qquad (3)$$

where $\dagger$ represents the Hermitian conjugate, $\sigma^0$ is the 2-by-2 identity matrix and $\sigma^1, \sigma^2, \sigma^3$ are the 2-by-2 three Pauli spin matrices, which are defined as

$$\sigma^1 = \begin{pmatrix} 1 & 0 \\ 0 & -1 \end{pmatrix}, \sigma^2 = \begin{pmatrix} 0 & 1 \\ 1 & 0 \end{pmatrix}, \sigma^3 = \begin{pmatrix} 0 & -i \\ i & 0 \end{pmatrix}, \qquad (4)$$

and

$$E(r) = \begin{pmatrix} E_x(r) \\ E_y(r) \end{pmatrix}. \qquad (5)$$

Hence,

$$S_n(r) = \sum_{a,b} \sigma^n_{ab} E_a^*(r) E_b(r), (a, b = x, y). \qquad (6)$$

The fluctuations of each SP around their average value can be defined as

$$\Delta S_n(r) = S_n(r) - <S_n(r)> \quad (n = 0, 1, 2, 3), \qquad (7)$$

where $S_n(r)$ is the SP pertaining to a single realization of the field at a specific spatial point and $<S_n(r)>$ denotes its ensemble average. Now ensemble averaging is replaced by spatial averaging by considering spatial stationary and ergodicity in space at the observation plane and $r_2 = r_1 + \Delta r$[22, 23]. The 4 x 4 matrix $C_{nm}(\Delta r)$ of two-point SPs correlations can be defined as follows:

$$C_{nm}(\Delta r) = <\Delta S_n(r_1) \Delta S_m(r_1 + \Delta r)> \quad (n, m = 0, 1, 2, 3) \qquad (8)$$

Using the Gaussian moment theorem, 4 x 4 Stokes fluctuations correlations matrix is transformed as

$$C_{nm}(\Delta r) = \sum_{a,b} \sum_{c,d} \sigma^n_{ab} \sigma^m_{cd} W_{ad}(\Delta r) W^*_{bc}(\Delta r), \quad (a, b, c, d = x, y), \qquad (9)$$

The terms $W_{ad}(\Delta r), W^*_{bc}(\Delta r)$ in Eq. (9) will be defined later.
The elements of the 4 x 4 complex polarization correlation matrix are evaluated from Eq. (9) by considering spatial averaging and these elements are.

$$\begin{pmatrix} C_{00}(\Delta r) & C_{01}(\Delta r) & C_{02}(\Delta r) & C_{03}(\Delta r) \\ C_{10}(\Delta r) & C_{11}(\Delta r) & C_{12}(\Delta r) & C_{13}(\Delta r) \\ C_{20}(\Delta r) & C_{21}(\Delta r) & C_{22}(\Delta r) & C_{23}(\Delta r) \\ C_{30}(\Delta r) & C_{31}(\Delta r) & C_{32}(\Delta r) & C_{33}(\Delta r) \end{pmatrix} \qquad (10)$$

The elements $C_{20}(\Delta r), C_{21}(\Delta r), C_{30}(\Delta r),$ and $C_{31}(\Delta r)$ of the matrix are represented as

$$C_{20}(\Delta r) = 2\, Re\, [W_{xx}(\Delta r) W^*_{yx}(\Delta r) + W_{yy}(\Delta r) W^*_{xy}(\Delta r)], \qquad (11)$$

$$C_{21}(\Delta r) = 2\, Re\, [W_{xx}(\Delta r) W^*_{yx}(\Delta r) - W_{yy}(\Delta r) W^*_{xy}(\Delta r)], \qquad (12)$$

$$C_{30}(\Delta r) = 2\, Im\, [W_{xx}(\Delta r) W^*_{yx}(\Delta r) + W_{yy}(\Delta r) W^*_{xy}(\Delta r)], \qquad (13)$$

$$C_{31}(\Delta r) = 2\, Im\, [W_{xx}(\Delta r) W^*_{yx}(\Delta r) - W_{yy}(\Delta r) W^*_{xy}(\Delta r)], \qquad (14)$$

The real and imaginary parts of the CPCF are evaluated by adding $C_{20}(\Delta r), C_{21}(\Delta r), C_{30}(\Delta r),$ and $C_{31}(\Delta r)$ elements.

$$C_{Re}(\Delta r) = C_{20}(\Delta r) + C_{21}(\Delta r), \tag{15}$$

$$C_{Im}(\Delta r) = C_{30}(\Delta r) + C_{31}(\Delta r), \tag{16}$$

Now, the CPCF is represented as

$$C(\Delta r) = C_{Re}(\Delta r) + i * C_{Im}(\Delta r), \tag{17}$$

$$C(\Delta r) = 4\, Re\left[W_{xx}(\Delta r)W_{yx}^*(\Delta r)\right] + i4\, Im\left[W_{xx}(\Delta r)W_{yx}^*(\Delta r)\right], \tag{18}$$

The correlation between two orthogonal polarization components $W_{yx}^*(\Delta r)$ is expressed as

$$W_{yx}^*(\Delta r) = \int [A_0(\hat{r})A\exp(i\Psi)]^* \exp\left[i\frac{2\pi}{\lambda f}\Delta r.\hat{r}\right] d\hat{r}, \tag{19}$$

where $[A_0(\hat{r})A\exp(i\Psi)]^*$ indicates the target source structure at the diffuser plane and f is the focal length of the Fourier transforming lens.

Equation (18) states that the complex amplitude of the target beam can be recovered from the speckle pattern. Now the recovered complex field can be decomposed to examine the OAM states of the incident target beam.

To demonstrate the application of the proposed technique in the measurement of the OAM spectrum, we considered the FOV as a target. For fractional values of $l$ the phase factor in the Eq. (1) can be represented as $\exp(il\varphi)$ where $\Psi = l\varphi$ and characterized in terms of Fourier series or superposition of all the integer OAM modes as[13].

$$\exp(il\varphi) = \sum_{n=-\infty}^{\infty} C_n(l)\exp(in\varphi). \tag{20}$$

where $C_n$ can be represented as

$$C_n(l) = \frac{\exp(il\pi)\sin(l\pi)}{\pi(l-n)},$$

where *n* is an integer number. Therefore, the light beam with fractional OAM modes could be deemed as multiplexing integer OAM beams with different weights.

## Experiment and Results

The proposed experimental setup for measuring the OAM spectrum is shown in Fig. 2.

A spatially filtered He-Ne laser light beam of wavelength 633 nm is attenuated with a neutral density filter (NDF) which reduces the unwanted power of the beam. The half-wave plate (HWP) is used to orient the incoming beam at $45^0$ with respect to the horizontal direction. The 50:50 beam splitter divides the $45^0$ polarized beam into two equal intensity beams. The beam transmitted from BS is used to illuminate a phase-only spatial light modulator (SLM) with a resolution of 1920 x 1080 and a pixel pitch of 8 μm (Pluto from Holoeye) working in the reflection mode. The SLM is used to load the FOV phase structure. This SLM modulates only the x-polarization component which is loaded with the FOV and the y-polarization component remains intact i.e. plane wave. The reflected beam from SLM is directed towards ground glass (GG) by a reflection from BS. The GG is introduced in the path of the beam as a scattering media. Further, the beam propagates through the GG and is randomly scattered, the GG scrambles the incident light and generates a speckle pattern. The random field from the GG is Fourier transformed by a lens (L) of focal length 150mm as described by Eq. (2).

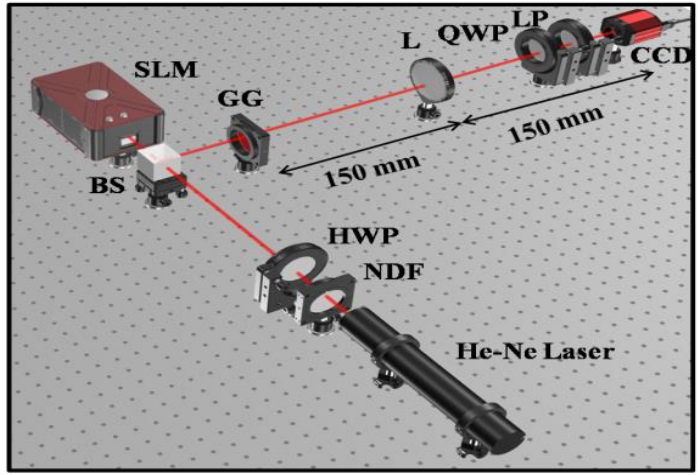

Figure 2. Sketch of the experimental set-up of the proposed technique. He-Ne Laser, NDF is a neutral density filter, HWP is a half-wave plate, BS is a beam splitter, SLM is a spatial light modulator, GG is ground glass, L is a lens, QWP is a quarter-wave plate, LP is a linear polarizer, CCD is a charge-coupled device.

The polarization states of the speckle pattern is characterized by measuring the SPs using a quarter-wave plate (QWP) and linear polarizer (LP) combination as shown in Fig. 2. The different combinations of QWP and LP are used to record intensity distribution of the speckle using a charge-coupled device (CCD) camera with a dynamic range 8-bit and resolution of 1280 x 1024 pixels and a pixel pitch of 4.65 micron [Thorlab model No. DCU224M]. The CCD captures the intensity pattern and the four SPs are determined from the captured speckle patterns by using the following equations[20].

$$S_0(r) = I(0°, 0°) + I(90°, 90°), \quad (21)$$

$$S_1(r) = I(0°, 0°) - I(90°, 90°), \quad (22)$$

$$S_2(r) = I(45°, 45°) - I(135°, 135°) \quad (23)$$

$$S_3(r) = I(0°, 45°) - I(0°, 135°), \quad (24)$$

where $I(\theta_q, \theta_p)$ is the intensity at the observation plane when the axes of the QWP and LP are at $\theta_q$ and $\theta_p$ respectively as measured from the horizontal direction.

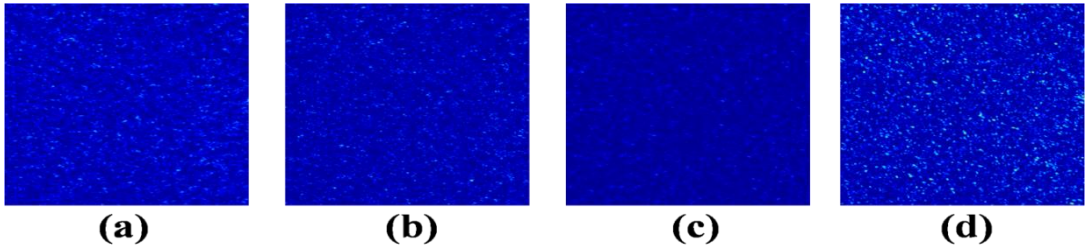

Figure 3. represent experimentally measured Stokes Parameters *(a) $S_0$ (b) $S_1$ (c) $S_2$ and (d) $S_3$.*

The experimentally measured SPs are used to evaluate the correlations of the fluctuations of the SPs. The experimentally measured SPs are used to extract the real and imaginary parts of the two-point CPCF using Eq. (15) and (16) respectively. The CPCF between two orthogonal polarization components is evaluated using Eq. (17). To evaluate the proposed technique, we performed computer simulation and experimental tests.

Simulation and experimental results of the amplitude and phase distribution of the incident FOV with $l = -0.5, 0.5,$ and $1.5$ are shown in Fig. 4 and Fig. 5. The CPCF encodes the complex FOV. Figs. 4,

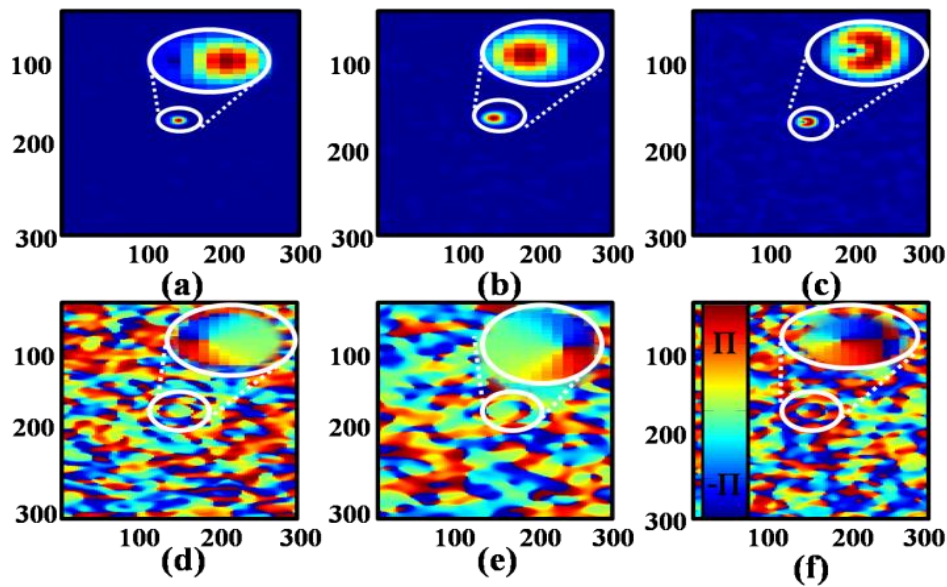

Figure 4. Simulation results; (a)-(c) represent amplitude distribution of CPCF for three different cases $l = -0.5, 0.5, 1.5$, and (d)-(f) are the corresponding phase distribution.

(a)-(c) show amplitude distribution and (d)-(f) show phase distribution of CPCF for $l = -0.5, 0.5, 1.5$. Figs. 5(a)-(c) and 5(d)-5(f) represent corresponding experimental results. Fig. 4, (a)-(c) represent simulated results of amplitude distribution, and experimental results in Fig 5, (a)-(c) reveal a dark line in the amplitude distribution of the CPCF. The OAM mode in one of the orthogonal polarization component of incoming light forms a dark line in the amplitude distribution of the CPCF. To quantitatively investigate the different integer OAM modes of FOV with $l = -0.5, 0.5, 1.5$, the spatial phase structure of the CPCF are shown in Figs. 4, (d)-(f) and 5, (d)-(f) for simulation and experimental results respectively. The spatial phase distribution of CPCF in Figs. 4 and Figs. 5 reveal the phase profile of FOV.

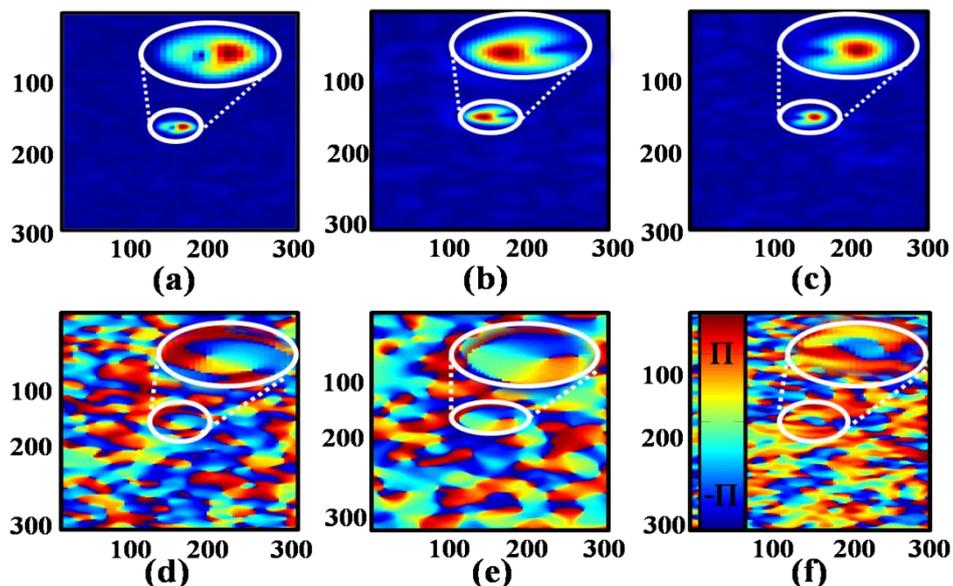

Figure 5. experiment results; (a)-(c) represent amplitude distribution of CPCF for three different cases $l = -0.5, 0.5, 1.5$, and (d)-(f) are the corresponding phase distribution.

## OAM Spectrum Analysis

The orthogonal projection method is used to decompose the different integer OAM modes in terms of the OAM power spectrum. To decompose the different integer OAM modes of the target FOV, an experimentally measured far-field FOV is projected onto spiral harmonics $e^{in\varphi}$, where $n$ is the topological charge. The complex coefficient $A_n$ is evaluated by integrating the recovered far-field FOV with respect to the azimuthal angle. The complex coefficient $A_n$ carries each OAM value as a function of the radial coordinates. Now, the OAM power spectrum of the beam is investigated by integrating the modulus square of $A_n$. Further, the OAM power spectrum is used to represents each OAM component in terms of azimuthal modes[20, 24].

The angular Fourier transform is applied over the recovered far-field FOV to evaluate the complex coefficient $A_l$.

$$A_n = \frac{1}{2\pi} \int_0^{2\pi} d\varphi e^{-in\varphi} C(\Delta r). \tag{25}$$

The complex coefficient is used to investigate the OAM power spectrum of the FOV beam by integrating $|A_n|^2$ along with the radial coordinates.

$$P(n) = \frac{1}{S} \int_0^\infty dr r |A_n|^2, \tag{26}$$

where, $S = \sum \int_0^\infty dr r |A_n|^2$ denotes beam power and $P(n)$ denotes the OAM power spectrum.

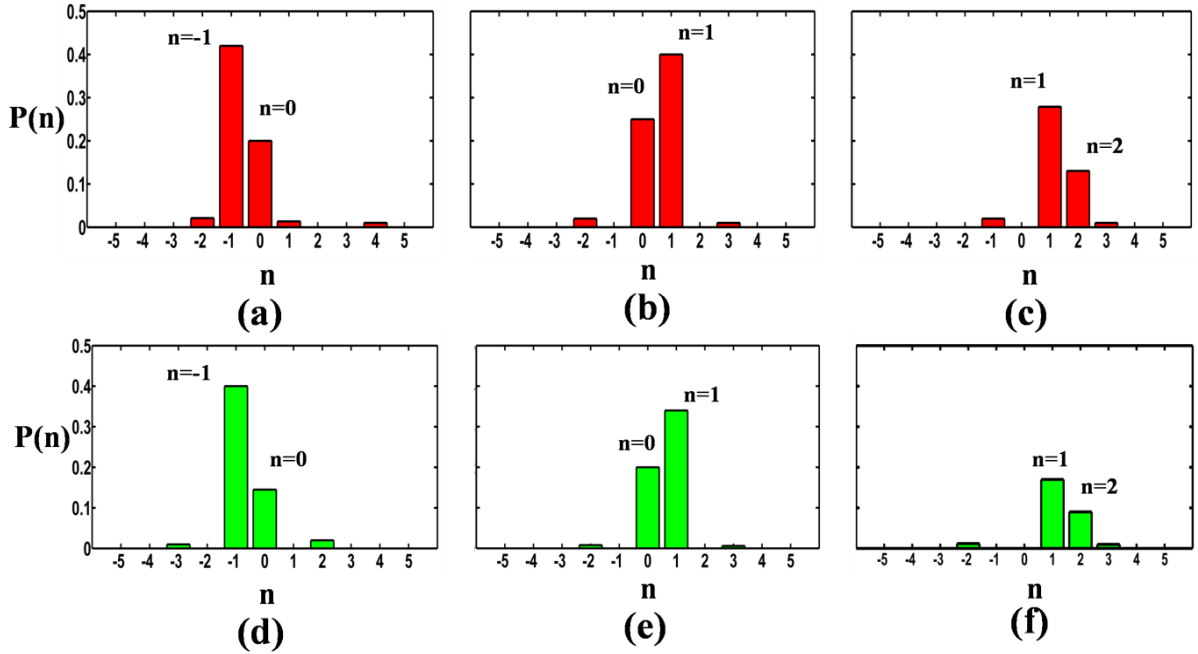

Figure 6. Panels (a)-(c) and (d)-(f) represent simulation and experimental results for the OAM distribution for three different OAM modes with $l = -0.5, 0.5, 1.5$.

Simulation and experimental results of the OAM distributions are shown in Fig. 6. Figs. 6 (a)-(c) and 6 (d)-(f) represent simulation and experimental results respectively. In Figs. 6 (a)-(c) and (d)-(f), the red and green color bars reveal the OAM distribution for three different OAM modes with $l = -0.5, 0.5, 1.5$, respectively. Fig 6(a) and 6(d) show OAM mode with $l = -0.5$ consisting of two integer modes $n = -1, 0$. Fig. 6(b) and 6(e) show OAM mode with $l = 0.5$ consisting of two integer OAM modes $n = 0, 1$, and Fig. 6(c) and 6(f) show OAM mode with $l = 1.5$ consisting of two integer OAM modes $n = 1, 2$.

## Conclusions

We have proposed, modeled, and experimentally demonstrated a quantitative technique to measure the OAM spectrum of the complex field obscured by random media. The developed theoretical framework is verified by simulation results and also tested by experimental demonstration. The applicability of the developed technique has been demonstrated experimentally to measure the OAM spectrum of the FOV for three different cases. The experimental results indicate that the proposed technique shows high flexibility and robustness. This technique is expected to play a crucial role in quantum-inspired imaging, cryptography, and optical communication.

## Acknowledgment
T. S. would like to acknowledge the University Grant Commission, India for financial support as Senior Research Fellowship. Supports from the Council of Scientific and Industrial Research (CSIR), India- Grant No 80 (0092) /20/EMR-II, Science and Engineering Research Board (SERB): CORE/2019/000026 are acknowledged in this work.

## Competing interests
The authors declare that there is no competing interest related to this paper.

## Author Contributions
R. K. S. and T. S. conceived the idea. T. S., R. P., and M. M. B. performed the experiments. All the authors analyzed the results, wrote and reviewed the manuscript.